\documentclass[fleqn,twoside]{article}

\newcommand{\bls}[1]{\renewcommand{\baselinestretch}{#1}}

\def\noi{\noindent}

\renewcommand{\subsection}{\@startsection{subsection}{2}{0pt}%
        {-3ex plus -1ex minus -.2ex}{1.4ex plus .2ex}%
        {\normalsize\bf\protect\raggedright}}

\renewcommand{\thesubsubsection}%
        {\arabic{section}.\arabic{subsection}.\arabic{subsubsection}.}
\makeatletter
\renewcommand{\@oddhead}{\raisebox{0pt}[\headheight][0pt]{%
   \vbox{\hbox to\textwidth{\rightmark \hfil \rm \thepage \strut}\hrule}}}
\renewcommand{\@evenhead}{\raisebox{0pt}[\headheight][0pt]{%
   \vbox{\hbox to\textwidth{\thepage \hfil \leftmark \strut}\hrule}}}
\makeatother
\newcommand{\heads}[2]{\markboth{\protect\small\it #1}{\protect\small\it #2}}

\newcommand{\Arthead}[5]{ \setcounter{page}{#4}\thispagestyle{empty}\noi
    \unitlength=1pt \begin{picture}(500,40)
        \put(0,58){\shortstack[l]{\small\it Gravitation \& Cosmology,
                        \small\rm Vol. #1 (#2), No. #3, pp. #4--#5\\
        \footnotesize \copyright \ #2 \ Russian Gravitational Society} }
    \end{picture}
	 }
\newcommand{\Title}[1]{\noi {\Large #1} \\}
\newcommand{\Author}[2]{\noi{\large\bf #1}\\[2ex]\noindent{\it #2}\\}

\newcommand{\Rec}[1]{\noi {\it Received #1} \\}

\newcommand{\Abstract}[1]{\vskip 2mm \begin{center}
        \parbox{16.4cm}{\small\noi #1} \end{center}\medskip}

\newcommand{\foom}[1]{\protect\footnotemark[#1]}

\newcommand{\email}[2]{\footnotetext[#1]{e-mail: #2}}

\def\nq{\hspace*{-1em}}
\def\nqq{\hspace*{-2em}}
\def\nhq{\hspace*{-0.5em}}

\def\cm{\hspace*{1cm}}
\def\inch{\hspace*{1in}}

\def\Jl#1#2{{\it #1\/} {\bf #2},\ }
\def\CQG#1 {\Jl{Clas. Qu. Grav.}{#1}}
\def\DAN#1 {\Jl{Dokl. AN SSSR}{#1}}
\def\GC#1 {\Jl{Grav. \& Cosmol.}{#1}}
\def\GRG#1 {\Jl{Gen. Rel. Grav.}{#1}}
\def\JETF#1 {\Jl{Zh. Eksp. Teor. Fiz.}{#1}}
\def\JMP#1 {\Jl{J. Math. Phys.}{#1}}
\def\NP#1 {\Jl{Nucl. Phys.}{#1}}
\def\PLA#1 {\Jl{Phys. Lett.}{#1A}}
\def\PLB#1 {\Jl{Phys. Lett.}{#1B}}
\def\PRD#1 {\Jl{Phys. Rev.}{D\ #1}}
\def\PRL#1 {\Jl{Phys. Rev. Lett.}{#1}}
\newcommand{\eqsection}{\makeatletter
	\@addtoreset{equation}{section}
	\renewcommand{\theequation}{\arabic{section}.\arabic{equation}}
	\makeatother}
\def\al{&\nhq}
\def\lal{&&\nqq {}}
\def\eq{Eq.\,}

\def\beq{\begin{equation}}
\def\eeq{\end{equation}}
\def\bear{\begin{eqnarray}}
\def\bearr{\begin{eqnarray} \lal}
\def\ear{\end{eqnarray}}
\def\earn{\nonumber \end{eqnarray}}
\def\nn{\nonumber\\ {}}

\def\nnn{\nonumber\\ \lal }

\def\yy{\\[5pt] {}}
\def\yyy{\\[5pt] \lal }
\def\eql{\al =\al}

\def\dst{\displaystyle}
\def\tst{\textstyle}
\def\fracd#1#2{{\dst\frac{#1}{#2}}}
\def\fract#1#2{{\tst\frac{#1}{#2}}}

\def\half{{\fract{1}{2}}}

\def\e{{\,\rm e}}
\def\d{\partial}

\newcommand*{\pdfq}[1]{\d^2_{#1}}
\newcommand*{\pdf}[1]{\d_{#1}}
\newcommand*{\pdi}[1]{\frac 1i \d_{#1}}

\heads  {Nikolay Popov and Petr Roshchin}
 	{A Geometric Model of Gravitating Mass Formation
	and Baryon Mass Spectrum}

\bls{0.96}
\begin{document}
\twocolumn[
\Arthead{7}{2001}{1 (25)}{1}{6}

\vspace*{-7mm}

\Title{A GEOMETRIC MODEL OF GRAVITATING MASS FORMATION \yy
	AND BARYON MASS SPECTRUM}

\Author{Nikolay Popov\foom 1 and Petr Roshchin}
{Computation Centre of the Russian Academy of Sciences}

\Rec{31 May 2000}

\Abstract
{The operator of elementary particle mass spectrum is
constructed on the basis of the postulate of a six-dimensional space-time
manifold having a Riemann-Cartan space structure, proceeding only from
its geometric properties. Special solutions found for this operator
well describe the mass spectrum of the known quark triplet and hence of some
baryon supermultiplets.}

\vspace*{-1.5mm}
] 

\email 1 {nnpopov@mail.ru}

\section{Introduction}

In \cite{Popov1998}, using a Riemann-Cartan six-dimensional manifold of signature $(- - - + + +)$, a purely geometric model has been successfully constructed, describing the gravitational field produced by a point source of time axis rotation in the temporal subspace. 
Thus one of the principal results of this study is that, according to the proposed geometric model, the gravitating mass concentrated at a given point of space originates due to rotation of the time axis associated with that point. 
It is noted that the gravitational field itself produced by such rotation coincides with the Schwarzschild field in the four-dimensional space-time submanifold, while the point mass turns out to be proportional to the squared angular velocity of time axis rotation. 
These results, based on new geometrical notions of space and time, are rather qualitative than quantitative and do not allow one to explain the existing laws governing the elementary particle mass spectrum. 
To solve such a problem, the formalism of quantum mechanics should be used.

If the above geometric model is sufficiently realistic, one would expect that the elementary particle mass values, for particles considered, as a first appoximation, as points, should represent eigenvalues of some invariant operator, uniquely induced by the metric of the space-time manifold constructed in \cite{Popov1998}. 
A significant part in the mass spectrum formation should be played, according to the geometric model, by the operator component that acts in the three-dimensional temporal subspace. 
The present study is concerned with solving this problem by establishing a strong relationship between the new notions of physical space-time structure and the properties of elementary particles.

\section{Statement of the problem}

Let a six-dimensional manifold $M^6$ be equipped with a Riemann-Cartan space structure, which is characterized by the metric $g_{ij}$ and the connection $\Delta_{ij}$. 
The tangential subspace at each point of $M^6$ is the pseudo-Euclidean space $R_{3,3}$ of signature ${(---+++)}$.
It should be noted that for the subsequent constructions it will be sufficient to know the metric $g$, which is equivalent to assigning to $M^6$ the structure of a pseudo-Riemannian space. 
However, proceeding in the spirit of \cite{Popov1998}, to which we shall have to refer in what follows, we prefer to leave everything as it is. Let  $ x_1, ... , x_6$ be pseudo-Euclidean coordinates in $M^6$, with respect to which the following requirements are satisfied:

\medskip\noi
1. The metric tensor field $g_{ij}(x)$ is continuously differentiable in the entire manifold $M^6$, except the line $x_1=x_2=x_3=x_5=x_6 = 0$, where the metric may be discontinuous.

\medskip\noi
2. The metric components $g_{ij} \; i,j=1,...,4$ are independent of $x_4, x_5, x_6$, while $ g_{ij}$, $i,j=5,6$ depend on $x_5, x_6$ only.

\medskip\noi
3. $g_{ij}=g_{ji}=0 $ \ for \ $i=1,2,3$ and $ j=4,5,6$.

\medskip\noi
4. The metric components $g_{ij}$ are invariant under orthogonal transformations of the coordinates $x_1,\ x_2,$  $x_3$ and, separately, $x_5,\ x_6$.

\medskip\noi
5. At infinity all the metric components tend to zero, except six, which have the following limiting values:
\[
g_{11}=g_{22}=g_{33}=-1 , \cm g_{44}=g_{55}= g_{66}=1 .
\]

Our task is to construct, on the manifold $M^6$ with a given metric $g$, the most general form of a self-adjoint differential operator of second order, $\Delta_{3,3}$, which possesses the following properties:

\medskip\noi
$1'.$ The operator $\Delta_{3,3}$ is invariant under the group of transformations $O(3)\times O(2)$;

\medskip\noi
$2'.$ $\Delta_{3,3}$ takes at infinity the following form:
\[
\Delta_{3,3} = \eta_{ij}\d_i\,\d_j\; ,
\]
where $\eta_{ij}$ is the pseudo-Euclidean metric in the space $R_{3,3}$ (here and henceforth $\d_i \equiv \d/\d_{x_i}$ and, for any $z$, $\d_z \equiv  \d/\d z$).
We must then investigate its spectrum and establish a direct relationship of this spectrum with the mass spectra of some elementary particles.

\section{Construction of the operator $ \bf{\Delta_{3,3}} $ }

Let us introduce the linear element
\[
 	ds={(g_{ij}\,dx_i\, dx_j)}^{1/2}
\]
Then the most general form of the squared linear element, according to \cite{Popov1998} and taking into account the requirements 2--5, can be represented as
\bear
	ds^2\eql F(r)dx_4^2 + M(\rho)(dx_5^2+dx_6^2)  \nnn\
		+ N(\rho)(x_5dx_5+x_6dx_6)^2 \nnn\
		- G(r)(dx_1^2+d\,x_2^2+d\,x_3^2)  \nnn \
	- H(r)(x_1d\,x_1+x_2d\,x_2+x_3\,dx_3)^2, 
\ear
where
\[
\rho=(x_5^2+x_6^2)^{1/2},\cm   r=(x_1^2+x_2^2+x_3^2)^{1/2}.
\]
The functions  $F,\,G,\,H,\,M$ and $ N $ have been found in \cite{Popov1998}, the first three of them coinciding with similar functions for the
Schwarzschild metric in the four-dimensional manifold \cite{Schwarzschild1916}. 
In the region $r > \varepsilon_2\;, \rho > \varepsilon_1$, where $\varepsilon_1, \varepsilon_2$ are arbitrary but small quantities, these functions have the form
\bear
	   F(r)\eql 1-\frac{a}{(r^3+a^3)^{1/3}}, \quad
 				G(r) = \frac{(r^3+a^3)^{2/3}}{r^2},
\nn
       H(r) \eql \frac{(r^3+a^3)^{-4/3} \, r^2}{1- a (r^3+a^3)^{-1/3}}
		       		- \frac{(r^3+a^3)^{2/3}}{r^4},
\nn
	M(\rho)\eql \frac{\rho^2+\beta^2}{\rho^2}, \quad
        N(\rho)=\frac{1}{\rho^2+\beta^2}-\frac{\rho^2+\beta^2}{\rho^4}, 
\ear
where $ a $ is the Schwarzschild radius and  $ \beta $ is a constant.
The most general form of a self-adjoint second-order differential operator which satisfies the properties $(1',2')$ is as follows:
\bearr
    \Delta_{3,3}=F(r)\d^2_{4} - \sum_{k=5,6} \pdi{x_k} M(\rho)\pdi{x_k}
\nnn
    -\Bigl(\pdi{x_5}x_5 + \pdi{x_6}x_6\Bigr)N(\rho)
			\Bigl(x_5\pdi{x_5} +  x_6 \pdi{x_6} \Bigr)
\nnn  \inch
     +\sum_{k=1,2,3} \pdi{x_k}\,G(r)\pdi{x_k}
\nnn                                                                 
     +\Bigl(\pdi{x_1}x_1+\pdi{x_2}x_2+\pdi{x_3}x_3\Bigr)H(r)
\nnn  \cm\cm \times
	\Bigl(x_1\pdi{x_1}+x_2\pdi{x_2}+x_3 \pdi{x_3}\Bigr),
\ear
provided that $ r > \varepsilon_2 \,,\quad \rho > \varepsilon_1 $.
Formally, the operator (3) is obtained from the squared linear element (1) by substituting the Hermitian operator $(1/i)(\d/\d x_j)$ for the differential $dx_j $ and appropriate regrouping of the operator cofactors. 
Before proceeding to determine the range of definition, $ D\,(\Delta_{3,3})$, of the operator (3), let us present it in new, partially spherical and partially cylindrical coordinates $ r,\,\theta,\,\varphi,\,\rho,\,\psi,\,t $ where $x_1=r\sin{\theta}\cos{\varphi}$, $ x_2=r\sin{\theta}\sin{\varphi}$, $x_3=r\cos{\theta}$, $x_4=t$,  $x_5=\rho\cos{\psi}$, $x_6=\rho\sin{\psi}$,
Then
\bearr
	x_5\d_{5}+x_6\d_{6} = \rho\pdf{\rho} \, ,
\nnn
	\d^2_{5} + \d^2_{6} = \pdfq{\rho}+\frac{1}{\rho}\pdf{\rho}+
			\frac{1}{{\rho}^2}\pdfq{\psi},
\nnn
	x_1\d_{1}+x_2\d_{2}+x_3\d_{3} = r\pdf{r} \, ,
\nnn
	\d^2_{1} + \d^2_{2}+ \d^2_{3} = \pdfq{r} + \frac{2}{r}\pdf{r} +
		\frac{1}{r^2}\Lambda(\theta,\varphi)\; ,
\earn
where
\[
\Lambda(\theta,\varphi) = \pdfq{\theta}+\frac{\cos\theta} {\sin\theta} \pdf{\theta} +\frac{1}{\sin^2\theta}\pdfq{\varphi},
\]
and the operator (3) will be represented in the new coordinates as follows:
\bear
	\Delta_{3,3}\eql F(r)\pdfq{t}+(M + {\rho}^2  N)
		\Bigl(\pdfq{\rho}+\frac{1}{\rho}\pdf{\rho}\Bigr)
\nnn
       + (M' + \rho^2 N' + \rho N)\,\pdf{\rho}+ \frac{M}{\rho^2}\pdfq{\psi}
\nnn \nqq \nq
      -(G + r^2 H) \Bigl(\pdfq{r} + \frac{2}{r}\pdf{r}\Bigr)
    -(G'+r^2 H')\pdf{r} - \frac{G}{r^2}\Lambda(\theta ,\varphi) .\nnn
\ear
The change from one coordinate system to another is achieved in a covariant manner.
The operator (4) may be written as a sum of two commutative operators,
\[
\Delta_{3,3} = \Delta^{(1)}_{3,3} + \Delta^{(2)}_{3,3} ,
\]
where
\bearr
	 \Delta^{(1)}_{3,3} = (M + {\rho}^2  N)\Bigl(\pdfq{\rho}+
				\frac{1}{\rho}\pdf{\rho}\Bigr)
\nnn \cm\cm
 	+\Bigl(M'+\rho^2 N'+\rho N \Bigr)\pdf{\rho} +
				\frac{M}{\rho^2}\pdfq{\psi},
\\ \lal
	\Delta^{(2)}_{3,3} = F(r)\pdfq{t} - (G +r^2  H)
				\Bigl(\pdfq{r}+\frac{2}{r}\pdf{r}\Bigr)
\nnn \cm\cm
	-\Bigl(G'+r^2\,H'\Bigr)\pdf{r}
			-\frac{G}{r^2} \Lambda(\theta,\varphi).
\ear
The range of definition $D (\Delta^{(1)}_{3,3})$ of the operator (5) consists of doubly differentiable, quadratically integrable functions of the parameters  $\rho,\ \psi$ relative to the measure $ d\mu_1 = \rho\, d\rho\, d\psi$\, which satisfy the condition
\beq
	\int_0^\infty  |g|^2 \rho^{-3} d\rho < \infty.
\eeq

The condition (7) is sufficient to ensure that
\[
{\Delta}^{(1)}_{3,3}\,g \in L^2 (R_2 , \mu_1 ) \qquad\mbox {for any}\qquad g\in D (\Delta^{(1)}_{3,3}).
\]
Taking into account that the operator $ \Delta^{(1)}_{3,3} $ in the form (5) is defined only for $ \rho>\varepsilon_1$, let us confine ourselves to considering the action of the operator $\Delta^{(1)}_{3,3}$ upon functions belonging to $ D(\Delta^{(1)}_{3,3})$ and defined in the domain
\[
R_2\setminus O_1,\quad\mbox{where }\quad O_1 = \{ \rho : \rho < \varepsilon_1 \}\, .
\]

The range of definition $D(\Delta^{(1)}_{3,3})$ of the operator (6) consists of twice differentiable and square-integrable functions of the parameters $ r, \theta, \varphi, t $ relative to the measure $d\mu_2 = r^2 \sin \theta\, dr\, d\theta\, d\varphi\, dt$, which
satisfy the condition
\beq
\int_0^\infty |g|^2 \, r^{-2} dr < \infty . 
\eeq
Integration in the parameter $ t $ is done in an arbitrary compact. 
The condition (8) is sufficient for the operator $ \Delta^{(2)}_{3,3} $ to be essentially self-adjoint. 
Since the operator  $\Delta^{(2)}_{3,3} $ in the form (6) is only defined at $ r>\varepsilon_2 $, let us  restrict ourselves to considering its action upon the functions constituting $D(\Delta^{(2)}_{3,3}) $ and defined in the domain
\[
R_{3,1}\setminus O_2 \, ,\quad \mbox{where} \; \ \
O_2 = \{ r : r <\varepsilon_2 \}\,.
\]

\section{Eigenvalues of the operator $\bf {\Delta^{(1)}_{3,3}} $}

With \eq (2), one can represent the operator (5) in the form
\bearr
     \Delta^{(1)}_{3,3} = \frac{\rho^2}{\rho^2 + \beta^2}\,
	\pdfq{\rho} + \Bigl(\frac{2\,\rho\,\beta^2}{(\rho^2+\beta^2)^2} +
        	 \frac{\rho^2 +\beta^2}{\rho^3}\Bigr)\pdf{\rho}
\nnn \inch
	+ \frac{\rho^2 + \beta^2}{\rho^4}\pdfq{\psi}\;.
\ear
We will only be interested in positive eigenvalues of the operator (9).
It can be noted that the essential spectrum of the formally self-adjoint operator (9) represents a negative semiaxis; therefore, all non-negative eigenvalues belong to a discrete spectrum. 
Accordingly, the problem of finding the eigenvalues reduces to that of solving the differential equation
\beq
     \Delta^{(1)}_{3,3} g = {\lambda}^2 g, \cm  g\in D(\Delta^{(1)}_{3,3}).
\eeq
We will seek the function $ g $ in (10) in the form
\beq                                                                  
   g(\rho,\psi)=\e^{im\psi} g(\rho),\quad m=0,\ \pm \half,\ \pm 1,\ldots.
\eeq
Substituting (11) into (10) and neglecting the terms containing the first derivative with coefficient $\beta/\rho$ in powers 2 and higher yields
\beq\nq\,
	\Bigl[\frac{{\rho}^2}{{\rho}^2 + {\beta}^2}\Bigl(\pdfq{\rho}
		+ \frac{1}{\rho}\pdf{\rho}\Bigr) -
    \frac{\rho^2 + \beta^2}{\rho^4}m^2\Bigr] g(\rho) = {\lambda}^2 g(\rho).
\eeq
Let us introduce the new variable  $x = \beta^2/\rho^2$,
$x \in (0, \beta^2/\varepsilon_1^2)$, then the relation (12) assumes the form
\bearr \nq\,
	\biggl[\frac{4 x^3}{{\beta^2}(x+1)}
		\Bigl(\pdfq{x}+\frac{1}{x}\pdf{x}\Bigr)
 			- \frac{x(x+1)}{{\beta}^2} m^2 \biggr] g(x)  
		= \lambda^2 g(x) ,                   \nnn
\ear
and the function $ g $ should be square-integrable in the region
$ (0, \beta^2/\varepsilon_1^2)$
over the measure $ d\mu = [\beta^2/(2x^2)] dx$, or, taking into account the condition (7), we arrive at the requirement
$ \dst\int_0^{\beta^2 \varepsilon_1^{-2}} |g|^2 \, dx< \infty $.
For $ x>0 $, we can write down \eq (13) in the form
\bear
   \biggl(\pdfq{x} + \frac{1}{x}\pdf{x} - \frac{m^2}{4} - \frac{m^2}{2x} -
	\frac{m^2 +{\beta}^2{\lambda}^2}{4x^2} \qquad \nn
		  -\frac{{\lambda}^2 {\beta}^2}{4x^3}\biggr) g(x) = 0. 
\ear
We will seek the function $ g(x) $ in (14) in the form $ g = x^{-1/2} v $, then for $v$ we get the equation
\beq    \nq\,
	v''-\frac{m^2}{4}v -\frac{m^2}{2}\frac{v}{x} - \frac{m^2 +
	\beta^2\lambda^2 -1}{4x^2}v-\frac{\beta^2\lambda^2}{4x^3} v = 0.
\eeq

\section{Finding a partial spectrum of $\bf{ \Delta^{(1)}_{3,3}} $}

Let as set $\lambda=0 $ in (15) and find the eigenfunctions and allowed values of the parameter $ m $ which correspond to the eigenvalue $ \lambda=0$ of the operator $ \Delta^{(1)}_{3,3} $. 
Consider the equation
\beq
v'' - \frac{m^2}{4}v - \frac{m^2}{2}\frac{v}{x} + \frac{1-m^2} {4x^2}v = 0.
\eeq
We will seek its solution in the form
\beq
v = \e^{-mx/2}\sum_{k=-1}^{\infty} C_k x^{r+1+k} \;.
\eeq
The lower limit of the sum in (17) is chosen on the basis of the requirement of quadratic integrability of the function $v\,x^{-1/2}$ near zero. 
Substituting (17) to (16) and denoting $y = \sum\limits_{k=-1}^{\infty}C_k x^{r+1+k}$, we arrive at the relation
\beq
	y'' - m y' -\frac{m^2}{2x}y + \frac{1-m^2}{4x^2}y = 0 .
\eeq
From (18) we obtain an infinite set of recurrent relations:
\bearr
	\Bigl[ (r+k)(r+k+1)  + \fract 14 (1-m^2) \Bigr] C_k
\nnn \qquad
   - m \Bigl(r+k+\frac{m}{2} \Bigr) C_{k-1} = 0, \qquad  k =0,1,2,... \;,
\earn
with a condition on $r$ in the form
\[
r=-\frac{B}{2}+\frac{m}{2}-\frac{1}{3}, \quad \mbox {where} \quad B=\frac{1}{3}\;.
\]
Hence,
\bear
	C_k =\frac{m\Bigl(m-\frac{B}{2}+k-\frac{1}{3}\Bigr)}{k\;(m+k)} 
 		C_{k-1}\; , \quad k = 0,1,... \;.
\ear
From (19) it follows that the sum $\sum\limits_{k=-1}^{\infty}C_{k+1} x^{r+1+k}$ should contain a finite number of terms to assure that the function $v$ be square integrable. 
If $m \ne 0$, then at some $m, k, B$ the equality
\beq
k - \frac{1}{3} + m -\frac{B}{2} = 0
\eeq
should hold. 
The parameters $\lambda\beta,\, m,\, k -\frac{1}{3},\, B$ admit the following physical interpretation: let $Q$ be the quark charge, $ B $ the baryon charge, $ I $ the isotopic spin, $I_3$ a projection of the isotopic spin, $S$ the particle strangeness, then
\[ \nq
S = -2\lambda\beta,\quad  Q=k-\fract{1}{3},\quad I = |m|,\ \ I_3=-|m|, ..., |m|.
\]
It can be noted that the idea to consider the isospin as a spin of a particle in the three-dimensional time-space was first advanced in \cite{Popov1997}. 
From (20) we obtain the following relation:
\beq
S = -2\,I_3-B+2\,Q                            
\eeq
for the case $ S = 0$. 
The relation (21) is nothing else but the Nishijima formula for the baryon supermultiplet. 
If $ S = 0 $ and $ m = 1/2$, then $ I_3 = \pm \half $, \  $ Q=- \frac{1}{3}\,,\frac{2}{3}\,,$ which corresponds to the quark doublet $ ( u,d )$.

Let us consider \eq (15) with $ m = 0 $, then we obtain the following relation:
\beq
v'' + \frac{1-\lambda^2\beta^2}{4x^2} \,v - \frac{\lambda^2\beta^2} {4x^3} \,v = 0 .
\eeq
We will seek its solution in the form
\beq
v = \e^{- \lambda\beta/\sqrt x} \, y .         
\eeq
Substituting (23) into (22), we come to the following set of equations for $y$:
\bear
y'' +\frac{1-\lambda^2\beta^2}{4 \;x^2} y = 0,\cm
y' - \frac{3}{4}\;\frac{y}{x} = 0 .
\ear
The system (24) has the solution $ y = cx^{3/4}$ only if $ \lambda \beta = \half$. 
This means that the Nishijima formula is true in the case $ \lambda\beta = \frac{1}{2} \:,\; I_3=0 \:,\;B= \frac{1}{3} \:,\; Q=-\frac{1}{3} \:,$ which corresponds to the quark $s$.

Thus we have found two eigenvalues $ 0 $ and $ 1/(2\beta)$ of the operator $ \Delta^{(1)}_{3,3} $, which correspond to the quark triplet $ (u,\ d,\ s)$.  
It can already be seen at this stage of the investigation that the $ s $ quark mass differs from the masses of $ u $ and $ d $ quarks.

\section{Finding a partial spectrum of $ \bf{\Delta^{(2)}_{3,3}} $}

With \eq (2), one can represent the operator (6) in the form
\bear
	\Delta^{(2)}_{3,3}\eql f_4\pdfq{t} -f_1 r^4 \pdfq{r}-\frac{f_2}{r^4}
 		\Lambda(\theta,\varphi)
\nnn \quad
	-\Bigl(\frac{2\,f_2}{r^3} +4\,f_1r^3 + f'_1\,r^4\Bigr)\pdf{r},
\ear
where $f_1 = G/r^4 + H/r^2,\  f_2 = r^2 G,\ f_4 = F$.

Accordingly, the problem of finding the eigenvalues reduces to that of solving the differential equation
\beq
 	\Delta^{(2)}_{3,3}\; g = \Delta^2\, g\;,
\eeq
where $g \in D (\Delta^{(2)}_{3,3})$. 
We will seek the function $g$ in (26) in the form
\beq
    g(t, r, \theta, \varphi) = G(t) R_{l,n}(r) Y_l(\theta, \varphi)
\eeq
where $Y_l$ is an eigenfunction of the operator $\Lambda (\theta,\varphi)$, corresponding to the eigenvalue  $-l\,(l+1)$, so that
\beq
	\Lambda (\theta,\varphi)\,Y_l(\theta, \varphi) = -l(l+1)Y_l(\theta,
		\varphi)\,;                                            
\eeq
$G(t)$ is an eigenfunction of the formally self-adjoint operator  $-\pdfq{t}$,
which has only an essential spectrum in the domain $ (0,\infty)$, so that
\beq
\pdfq{t}\,G(t) = \alpha^2\,G(t).             
\eeq
Taking into account the relations (25), (27), (28), (29), one can represent \eq (26) in the form
\bear
	\Bigl[f_4 \alpha^2 - f_1r^4\pdfq{r} - \Bigl(\frac
		{2\,f_2}{r^3}+4 \,f_1r^3    + f_1'r^4\Bigl)\pdf{r} \quad
\nn
	+\frac{f_2}{r^4}l\,(l+1)\Bigl] R_{l,n} =\Delta^{2}R_{l,n}.
\ear

The essential spectrum of the formally self-adjoint operator in the left-hand side of \eq (30) is represented by the range $ (\alpha^2 \,,\infty) $, therefore all non-negative eigenvalues of the operator $\Delta^{(2)}_{3,3}$ restricted above by the value $ \alpha^2 $ are discrete. 
From the condition $ f_1\,r^4 \ne 0 $ for $ r \in (\varepsilon_2,\infty) $ it follows:
\bear
  \biggl[\pdfq{r}+\Bigl(\frac{4}{r}+\frac{f_1'}{f_1}+2\frac{f_2}{r^7f_1}
		\Bigr)\pdf{r}- \frac{f_4}{f_1r^4}\alpha^2  \qquad \ \
\nn
	-\frac{f_2}{f_1r^8}l(l+1)
		+\frac{\Delta^2} {f_1r^4} \biggr]\,R_{l,n}=0.
\ear

We will seek its solution in the form
\beq
R_{l,n} = \exp \biggl[-\frac{1}{2}\int\Bigl(\frac{4}{r} + \frac{f_1'}{f_1} +
\frac{2\,f_2}{r^7 \, f_1}\Bigr)\, dr\biggr] v_{l,n}.           
\eeq
After substitution of (32), \eq (31) may be rewritten in the form
\beq
	v_{l,n}'' - \Bigl(\fract{1}{4} P^2 + \half P' +           
			Q \Bigr) v_{l,n} = 0,
\eeq
where \cite{Stepanov1958}
\[ \nq
   P = \frac{4}{r} {+} \frac{f_1'}{f_1} {+} \frac{2\,f_2}{r^7 f_1}; \
   Q = \frac{f_4}{f_1\,r^4}\alpha^2 + \frac{f_2}{f_1\,r^8}l(l{+}1) -
			\frac{\Delta^2} {f_1\, r^4}.
\]
Taking into account that
\bearr \nq
  \frac{1}{f_1 r^4}\sim 1 - \frac{a}{r}+ o\Bigl(\frac{a^2}{r^2}\Bigr),\
	\frac{f_4}{f_1r^4}\sim 1-\frac{2a}{r}+\frac{a^2}{r^2}
			+o\Bigl(\frac{a^2}{r^2}\Bigr),
\nnn
    \frac{f_2}{f_1r^6}\sim 1 - \frac{a}{r} + o\Bigl(\frac{a^2}{r^2}\Bigr),
    \quad
    \frac{1}{4}P^2 + \frac{1}{2}P' \sim  o\Bigl(\frac{a^2}{r^2}\Bigr)
\earn
and neglecting the terms containing $\frac{a}{r}$ in powers 3 and higher yields
\bear
	v_{l,n}''- (\alpha^2-\Delta^2 )v_{l,n} +\frac{a}{r} (2\,\alpha^2
		-\Delta^2)v_{l,n}   \qquad
\nn
	-\frac{a^2\alpha^2+l(l+1)}{r^2}\, v_{l,n} =0 .
\ear
We will seek the solution of \eq (34) in the form
\beq
	v = \exp [-(\alpha^2 - \Delta^2)^{1/2}r] y_{l,n},          
\eeq
where  $ y_{l,n} = \sum\limits_{k=0}^\infty C_k r^{\nu+k} $.
Substituting (35) into (34), we come to the following relation for $y_{l,n}$:
\bear
      y_{l,n}''-2\,(\alpha^2 - \Delta^2)^{1/2} y_{l,n}'+
		(2\alpha^2 - \Delta^2)\frac{a}{r}\, y_{l,n}
\nn
	-\frac{l(l+1) + a^2 \alpha^2 }{r^2} \,y_{l,n} =  0 \;.
\ear

\eq (36) gives an infinite set of recurrent relations
\bearr
	[(\nu+k+1)(\nu + k)-l(l+1)-a^2 \alpha^2\,)] \,C_{k+1}
\nnn \quad
	- [2(\alpha^2 - \Delta^2)^{1/2} (\nu + k)-
		(2\alpha^2 - \Delta^2) a ] C_k = 0 ,
\earn
where $k =0,1,2,\ldots$ and
\beq
	\nu = \half \pm \half \sqrt{1+4\,l(l+1)+4\,a^2\alpha^2}.
\eeq
Hence,
\beq  \nhq
	C_{k+1} = \frac{2(\alpha^2{-}\Delta^2)^{1/2}(\nu{+}k)
					-(2\alpha^2 -\Delta^2)a}    
	{(\nu+k+1)(\nu+k)-l(l+1)-a^2\alpha^2} C_k.
\eeq
From (38) it follows that the sum $ \sum\limits_{k=0}^\infty C_k\,r^{\nu+k} $ should contain a finite number of terms to assure that the function $ v_{l,n}$ be square-integrable in the region $(\varepsilon_2, \infty) $.
Thus at some $ k=n$ the equality
\beq
	2(\alpha^2-\Delta^2)^{1/2}(\nu+n)-2\Bigl(\alpha^2 -\frac
			{\Delta^2}{2}\Bigr) a = 0 .
\eeq
should hold, and for all $ k < n $ the inequality
\[
	(\nu+k+1)(\nu+k)-l(l+1)-a^2\alpha^2 \ne 0
\]
is valid. 
Solving \eq (39) for $ \Delta^2 $, we obtain
\bearr
    \Delta_{1,2}^2 = 2\biggl[\alpha^2-\Bigl(\frac{\nu+n}{a}\Bigr)^2\biggr]
	\pm 2\biggl[\alpha^2 -\Bigr(\frac{\nu+n}{a}\Bigr)^2\biggr]
\nnn \inch \times
    \biggl[1 - \frac{\alpha^2}{\alpha^2 - [(\nu+n)/a]^2}\biggr]^{1/2} . 
\ear
From (40) it follows
\beq
	\Bigl(\frac{\nu+n}{a}\Bigr)^2 >\alpha^2                 
\eeq
because $1 - \fracd{\alpha^2}{\alpha^2-[(\nu + n)/a]^2} \ge  0$.
Hence from (37), (40) and (41) we obtain for $\alpha^2$ and $\Delta^2$ the following relations:
\[
\alpha^2 = \frac{\nu(\nu-1)-l(l+1)}{a^2}, \quad
\Delta^2 = \frac{2}{a^2} A\biggl[\frac{|\nu+n|}{A^{1/2}}-1\biggr],
\]
where $A = (\nu + n)^2 + l(l+1) - \nu (\nu-1)$, $l$ is the spin moment of a particle, while the physical meaning of the quantum parameters $\nu$ and $n$ remains obscure.

\section{Partial spectrum of $\bf{\Delta_{3,3}}$ and the mass spectrum of the quark triplet }

The obtained partial spectra of the operators $\Delta_{3,3}^{(1)}$ and $\Delta_{3,3}^{(2)}$ lead to a partial spectrum of the operator $\Delta_{3,3}$:
\[
\Delta_{3,3}g = (\lambda^2 + \Delta^2) g ,
\]
or in an expanded form
\bearr
     \Delta_{3,3}\, g = \biggl[\Bigl(\frac{2Q - B -2I_3}{\beta}\Bigr)^2
		+\frac{2}{a^2} A \Bigl(\frac{\nu + n}{A^{1/2}}
			- 1\Bigr)\biggr] g.
\nnn
\ear
If $Q = -\frac{1}{3},\ \frac{2}{3},\ B = \frac{1}{3},\ I_3 = -\frac{1}{2},\ \frac{1}{2},\ 0,\  l=\frac{1}{2}$, then this set of quantum characteristics
corresponds to the quark triplet $ d,\ u,\ s$. 
From (42), for the mass doublet of the quarks $ (d, u) $ and the quark $ s $ we get the following relations:
\bearr
	m_{d,u} = \biggl \{\frac{2}{a^2}
		\biggl[(\nu+n)^2-\nu(n-1)+\frac{3}{4}\biggr]
\nnn \cm
	\times \biggl[
      \Bigl(1 - \frac{\nu(\nu-1) - 3/4}{(\nu+n)^2}\Bigr)^{-1/2}-1 \biggr]
      				\biggr\}^{1/2},
\yyy
	m_s = (\beta^{-2} + m_{d,u}^2)^{1/2} .
\ear

We note that the relations (43) and (44) are given in provisional units
of measurement $ h= 1$, $c=1$, $e=1$. 
Knowing the mass spectrum of the quark triplet, we can readily find the mass spectrum of, e.g., the baryon decuplet, ignoring the spin-spin and spin-orbital interactions. 
The result satisfactorily describes the nature of the supermultiplet spectrum.

\section {Conclusion}

The very attempt to relate the geometric properties of the space-time continuum to the properties of elementary particles, in particular, to the quark mass spectrum appears, judging from the results that have been obtained, to be promising. 
Apparently, the first theoretical substantiation of the Nishijima formula for the quark triplet $ (u,\ d,\ s) $ on the basis of the geometrical properties of space and time is not coincidental.

The main achievement of the suggested work is apparently the construction of an unlimited self-adjoined operator, possessing a discrete spectrum adequately describing the quark triplet mass spectrum.

\end{document}